\documentclass[conference]{IEEEtran}
\IEEEoverridecommandlockouts

\usepackage[bookmarks=false]{hyperref}

\newcommand{\orcid}[1]{\href{https://orcid.org/#1}{\textcolor[HTML]{A6CE39}{\aiOrcid}}}

\def\BibTeX{{\rm B\kern-.05em{\sc i\kern-.025em b}\kern-.08em
    T\kern-.1667em\lower.7ex\hbox{E}\kern-.125emX}}

\begin{document}

\title{A perspective to navigate the National Laboratory environment for RSE career growth
}
\author{\IEEEauthorblockN{William F. Godoy}
\IEEEauthorblockA{ Computer Science and Mathematics Division \\
Oak Ridge National Laboratory \\
Oak Ridge, TN, USA \\
godoywf@ornl.gov}
}

\maketitle

\begin{abstract}
This paper shares a perspective for the research software engineering (RSE) community to navigate the National Laboratory landscape. The RSE role is a recent concept that led to organizational challenges to place and evaluate their impact, costs and benefits. The premise is that RSEs are a natural fit into the current landscape and can use traditional career growth strategies in science: publications, community engagements and proposals. Projects funding RSEs can benefit from this synergy and be inclusive on traditional activities. Still, a great deal of introspection is needed to close gaps between the rapidly evolving RSE landscape and the well-established communication patterns in science.
This perspective is built upon interactions in industry, academia and government in high-performance computing (HPC) environments. The goal is to contribute to the conversation around RSE career growth and understand their return on investment for scientific projects and sponsors.
\end{abstract}


\section{Introduction}
While the term ``research software engineer" (RSE) is relatively new, the role has existed for decades~\cite{baxter2012research}. In recent years, important conversations around scientific software stewardship\footnote{\url{https://github.com/tgamblin/ascr-software-stewardship-rfi-responses}} and cultural challenges~\cite{8565942} elevate the value of RSE roles.
The United States Department of Energy (DOE) has a rich history in funding high-performance computing (HPC) software projects to accomplish its scientific mission~\cite{osti_6711815,doi:10.1177/1094342010391989}. Future trends present a great opportunity to rethink codesign~\cite{osti_1822198} and HPC software~\cite{osti_1471119} as a more dynamic and multidisciplinary endeavor~\cite{Carver2008}. RSEs are at the core of the DOE mission as stewards of the software that allow for scientific discovery.

The present document lists aspects of well-established channels in the scientific community that RSEs can adopt towards a sustainable career growth path. These views come from collected personal experiences and observations in HPC environments in industry, government and academia that could be extended to other scientific communities with similar challenges.

\section{Career Growth Aspects}
The list presented in this section comes up from personal interactions with members of the scientific community when discussing RSE roles. They provide a simple context on how RSEs' careers can develop within a scientific setting.

\paragraph{Define your science} RSE work can be varied and, many times, not directly related to novel scientific discovery: e.g. addressing bugs, consultant work, tracking issues, etc. Nevertheless, the answer to the question that comes up frequently: \textit{what exactly is your science?} should be effectively communicated. While RSEs with closer expertise in a scientific domain or those who have had these discussions are able to answer this question, others may start thinking about answers in terms of value. Use the scientific method by providing empirical evidence for how RSE work impacts the science-funded initiative. The latter requires a conversation with stakeholders to figure out what needs to be measured that is important to them.

\paragraph{The importance of publications} peer-reviewed scientific publications are a cornerstone of science. Scholarly work is how careers are built and it is the language used by scientists. Metrics such as the h-index measure the impact of individuals on the scientific community~\cite{bornmann2007we}.
RSEs have publication venues that provide citable sources and professional societies, such as US-RSE\footnote{\url{https://us-rse.org/}} to disseminate their work effectively. Project stakeholders should consider prioritizing RSE involvement in publications and proposals following ethical, organizational and national security guidelines with the goal of retaining valuable contributors. This becomes even more relevant as much of the software produced in government organizations might not fall under the free and open source (FOSS) category.
Using this very same language will ease the communication of the impact of RSE activities in the broader scientific community.

\paragraph{Scientific software as specialized equipment}
scientific software can have little resemblance to products in the broader software landscape. Therefore, the role might be closer to those building ``specialized equipment" in novel scientific facilities, this is even more palpable in HPC and experimental ans observational science (EOS). Scientific project stakeholders may establish a similar synergy that understands and values the complexities of RSE and traditional science activities. At the same time, project stakeholders should consider that methods applied in the broader ``commodity-like" software landscape might incur in extra overheads for scientific software ({\it e.g.} agile management, enterprise software solutions, paradigms, etc.). Their real cost/benefit must still be proven scientifically with empirical evidence, ideally in peer-review publications. Scientific software is messy and unpredictable by definition, requirements are not always known at development time, so a great deal of flexibility is needed to achieve scientific discovery goals.

\paragraph{Develop community} this is a crucial aspect as one of the important metrics of science is how your community perceives your work and participation. External recognition in professional societies and organizations ease the path for internal career growth. Prioritizing these aspects pays off as it builds trust in the RSE community. Identify key members in your community that can recommend your work. Value previous and new contributors from multiple disciplines that enable RSE work. 
While promote and mentor new members of your community through your network. Overall, careers are built through the synergies of enabling people, not always by particular technologies.
Build upon community-based consensus decisions rather than those based on an appeal to authority by a few individuals to align with science endeavors.
Work side-by-side, not head-to-head, including and listening to all stakeholders, regardless of hierarchical structures.
Software is a rapidly changing environment, so early advisory activities from all stakeholders is crucial for decision making.

\paragraph{Identify critical paths} RSE tasks can carry significant overhead to project goals that do not necessarily lead to career growth. Thus, RSEs may prioritize activities with a higher return on investment that will propagate through the ``critical path" of a project, organization, sponsor, and your community. Focus on reportable activities that go beyond one or two layers across these paths: papers, highlights, tutorials, meeting organizations, mentoring, awards, proposals etc. Observe the career evolution of more senior members in your organization and ask related questions, so RSEs can establish similar conversations with those stakeholders that are more likely to promote their community members. Identify and minimize engagements that are not on the critical path for career growth.

\paragraph{Scientific impact and software quality trade-offs} Software quality is not free. By embracing imperfection and focusing on the value of the return in terms of scientific outputs, RSEs can focus on career growth through diversifying their portfolio of investments beyond software specific activities. Not all software quality activities lead to high research impact so understanding the potential cost/benefit is important. In addition, since not all requirements are always known at development time, a high investment in quality might in fact be counterproductive.

\paragraph{Promote your work along with the software} RSEs may be dedicated to particular products or technologies. Ultimately, projects are funded for scientific discovery so sustainability aspects always come as a question for potential adopters. Be ready to answer questions like: ``what happens after funding ends?", or ``who supports this software in the long-term?". People behind a product are the ones building the trust among stakeholders.

\paragraph{Embrace senior roles} RSE work in early career stages is heavily defined by development related activities. As professionals advance in their career the impact on others become more relevant. Software is a rapid evolving landscape, so learning from every generation that grew up with different technologies is extremely beneficial. Therefore, embracing tasks beyond development such as, mentoring, learning and promoting other RSEs is required for career growth.

\section{Conclusions}
The positions presented on this paper argue that RSEs are a natural fit in and an important part of the codesign process as scientific projects become more multidisciplinary in nature. RSEs can benefit from traditional scientific metrics: papers, service, and proposals, by building a diversified portfolio of activities that impacts their scientific communities.
For the RSE role to be seen as the new wave of ``a discipline on its own", the scientific community should do a great deal of introspection and learn from the ``new" science and engineering disciplines of the past that established a scientific framework for successful career paths for its members.
Ultimately, research software engineering is a human activity that would not be possible without rewarding its stakeholders.
\bibliographystyle{IEEEtran}
\bibliography{IEEEabrv,paper.bib}

\section*{Acknowledgement}
This manuscript has been authored by UT-Battelle, LLC, under contract DE-AC05-00OR22725 with the US Department of Energy (DOE). The publisher acknowledges the US government license to provide public access under the DOE Public Access Plan (\url{https://energy.gov/downloads/doe-public-access-plan})

\end{document}